\documentstyle[12pt]{article}
\newcommand{\be}{\begin{equation}}
\newcommand{\ee}{\end{equation}}

\begin{document}
\begin{titlepage}

\begin{flushright}
UUITP-20/94
\end{flushright}
\vspace{20 mm}

\begin{center}
{\huge String Theory, Black Holes and Klein's Lemma\footnote[1]{Talk given at
the Oskar Klein Centenery Symposium, Stockholm, September 19-21, 1994.}}
\end{center}

\vspace{10 mm}

\begin{center}
Ulf H. Danielsson\\
Institute for Theoretical Physics, Uppsala University, Uppsala, Sweden
\end{center}

\vspace{2cm}

\begin{center}
{\large Abstract}
\end{center}
In this lecture, I will discuss the connection between entropy, black holes and
the measurement problem of quantum mechanics. I will show how the ideas of
Oskar Klein play an important role in a possible solution of the information
paradox: black hole complementarity.

\vspace{2cm}
\begin{flushleft}
UUITP-20/94 \\
November 1994
\end{flushleft}

\vspace{2cm}

 \end{titlepage}
\newpage

\section{Introduction}

The problem that I want to discuss in this lecture has ancient roots. Through
the ages it has taken different guises. A  hundred years ago many asked the
question: ``What is the fundamental meaning of thermodynamics?'' Today, as we
will see, some wonder whether the unitarity of quantum mechanics is broken.
Oskar Klein also had his thoughts on the subject, and this has been the
motivation behind my lecture. His famous lemma, which I will come back to,
contains ideas of extreme importance to my main theme.

A centerpiece in these discussions is the second law of thermodynamics.
Although formulated in the 19th century, its equivalent, described in different
words can be found much earlier. In fact, we need to go back to 100 B.C. when
Lucretius, in his ``De Rerum Natura'' (``On the Nature of Things''), wrote:

\begin{flushleft}
{\it ``Again, perceivest not \linebreak
How stones are also conquered by Time?- \linebreak
Not how the lofty towers ruin down, \linebreak
And boulders crumble?- Not how shrines of gods \linebreak
And idols crack outworn?- Nor how indeed \linebreak
The holy Influence hath yet no power \linebreak
There to postpone the Terminals of Fate, \linebreak
Or headway make 'gainst Nature's fixed decrees?''}
\end{flushleft}

Clearly, the idea that the decay of the worlds is a law of nature is not new.
Even the idea that
this change may be the cause of time can be found a long way back in history.
Indeed, in 400 A.D. Augustine wrote in his ``Confessions'':

\begin{flushleft}
{\it ``In you my mind , I say, I measure times. What I measure is the effect,
itself present, which passing things create in you, and which remains when they
have passed away. In measuring times, I do not measure the things which passed
in the act of creating that effect, but rather the effect itself. Therefore
either times are that effect, or I do not measure times at all.''}
\end{flushleft}

But it was not until Ludvig
Boltzmann in the 19th century, that these ideas could be put on a more firm
footing through the
discovery of statistical mechanics. The question of whether Boltzmann succeeded
in giving a complete account of the second law is a subtle one. Furthermore, as
we will see, the need for contemplating these questions can be found in
many unexpected areas of physics. It relates to issues like the
arrow of time, the measurement problem of quantum mechanics, and,
which is the main topic of my talk, to the black-hole information
paradox.

The precise question that I want to address is: ``Is thermodynamics
always `just' statistical mechanics, or is it, sometimes, something
more?". This question was discussed intensely by Boltzmann and his opponents.
An anti-atomist like Ernst Mach (perhaps better known for
his thoughts on relativity) certainly thought there were something more to
thermodynamics than statistical mechanics. It was
considered by Mach and his followers
to be a great loss if the second law of thermodynamics were
to be explained by pure microphysics. Clearly they were mistaken in their
views. Atoms do exist and thermodynamics is transcended by statistical
mechanics. But is this always true? Could there be situations where,
perhaps, there is no microscopical description?

Modern physics can supply a candidate; evaporating black holes.
It was discovered in the seventies \cite{hawk1} that black holes radiate due to
quantum effects. Hawking noted that a temperature,
\be
T= \frac{\hbar c^{3}}{8\pi GM} ,
\ee
could be associated with the radiation. Furthermore, the black holes could also
be assigned an entropy \cite{bek}  given by
\be
S=\frac{4\pi kGM^2}{c\hbar }=\frac{kc^{3}A}{4G\hbar}.
\ee
$M$ denotes the mass of the black hole and $A$ the area of the event horizon.
It was soon realized that the phenomenon of Hawking radiation leads to a
surprising paradox. Let me briefly explain why this is so.

When a black hole is formed, information about from what it was formed is
hidden from view behind the event horizon. This is no mystery. Even if the
information is not accessible to us, we can still assume that it is safely
stored inside the black hole. Eventually, it seems, the black hole will
completely evaporate due to the Hawking radiation. The only thing left will be
the radiation itself. Will this radiation carry a memory of what formed the
black hole? According to Hawking's original calculations, the answer is no. If
this is so, the evaporation process has destroyed information and produced
``true'' entropy. After all, since the black hole has disappeared, we can no
longer claim that information, although inaccessible, still exists behind an
event horizon.
Formulated in the language of quantum mechanics it means that unitarity is
broken. We loose the ability to predict, given an initial state, the precise
final state. We can only give probabilities for various wave functions. In
other words, the Schr\"{o}dinger equation no longer works \cite{hawk2}.

This is a serious problem. If the conclusion is correct the consequences are
far reaching. Not only would non-unitary processes be important for exotic
phenomena like macroscopic black hole evaporation, they would also have
consequences in more or less daily life. Virtual black hole formation and
evaporation would spoil unitarity even when there are no real black holes
around.  Clearly one must carefully explore possible loopholes in the argument.

I will consider three different ways of attacking the problem in
the black hole context. I will give a personal review of
the progress made and which are the implications, if any.

First I will consider the problem from a mathematical model point
of view. That is, investigate the possibility of finding a model of an
evaporating quantum black-hole where the consequences can be calculated!
Clearly this will meet with only partial success, hence
I will proceed by considering a less rigorous approach. One can try to {\it
invent}
scenarios which in a logical way provide some solution of the
problem. Whether these scenarios are realized in a model will be
left for future work. Finally, one can try to imagine possible
uses for true entropy. Do we need it in any way? Does it help
explaining other phenomena? Does it enrich physics? In many ways, the
discussion today resembles the one held a hundred years ago, but we have yet to
see the outcome this time around.

But first we must recall some basic thermodynamics. This will be
necessary for the following discussion, and we will also see that
Oskar Klein plays an important role.

\section{Entropy, the Arrow of Time and All That}

The main contribution of Boltzmann was his discovery of a microscopic
definition of entropy. He found that
\be
S= - k \int \rho \log \rho     ,
\ee
where $\rho$ is the probability density on phase space. It follows from this
definition that the entropy is larger if $\rho$ is smeared out. Think of a gas.
There are fewer ways to distribute the molecules in the corner of a container
than in the whole container. Given a macroscopic state, the entropy is higher
if there are more microscopic realizations.

But how can we understand the approach to thermal equilibrium? Why is the
entropy always increasing? The microscopic definition of entropy given by
Boltzmann does not, by itself, explain the second law of thermodynamics. On the
contrary, the time evolution of $\rho$ is that of an incompressible fluid, and
the entropy remains constant. To improve on this situation,
Boltzmann considered a gas of {\it colliding particles}. The collisions are
described by his famous {\it Sto$\beta$zahlansatz}. Initially the positions and
the velocities of the particles are assumed to be known with some accuracy. As
the gas evolves and the particles collide  some of the information about
positions and velocities is transferred into information about correlations of
the particles.  Thanks to the Liouville theorem the phase-space volume must
remain constant. But  in practise we do not care about correlations when we
study a gas of particles. If we ignore the correlations, the phase-space volume
will grow and hence the entropy increase.
More precisely, instead of using the full $6N$ dimensional phase space (called
$\Gamma$-space by Boltzmann), where the system of N particles corresponds to a
single point, he considered a $6$ dimensional phase space with $N$ points
(called $\mu$-space). While the entropy remains constant using $\rho
_{\Gamma}$, it will increase if $\rho _{\mu}$ is used.

If one is interested in crystals and other regular systems, however,  it is not
appropriate to ignore correlations.  In these cases we must follow Gibbs, and
work with the $\Gamma$-space directly. Then, again, how can entropy increase?
Luckily, even the entropy of Gibbs can be made to increase after  some
convenient coarse-graining. Let us assume that  we can study the phase space
only with some finite resolution, and that the entropy is computed after the
phase space distribution has been smoothed. The smoothing will enlarge the
phase space volume, hence the entropy. The difference will be insignificant if
there is no structure at the resolution scale.
If there is a lot of structure on small scales however, the change will be
considerable. Therefore, if a smooth phase space volume evolves through time
into a complicated shape with a lot of structure on small scales it follows
that entropy will increase.
Entropy increase is in general related to information loss of the above type.
Relevant information is transformed into irrelevant information as far as the
macroscopic description is concerned.

Unfortunately the argument can be run backwards. A typical microscopic
representative of a given macroscopic state will evolve towards larger entropy
in either direction of time. We are forced to conclude that entropy will not
only be larger in the future, it must also have been larger in the past.  But
this does not agree with experiment! We have therefore failed to derive the
arrow of time. The solution of this paradox is to realize that the {\it real}
microscopic representatives are {\it not} typical as far as backwards evolution
is concerned. The world seems to be the product of very special initial
conditions. On the other hand, entropy increase also means that there are no
{\it final} conditions. These would appear as inexplicable conspiracies.
Boltzmann were aware of theses problems. He imagined that the initial
conditions were a product of some very unlikely chance fluctuations. In the
long run, there would be no preferred direction of time. A nice discussion of
these and other related issues can be found in [9].

So far we have only discussed classical physics. How do we formulate these
questions in a quantum framework?  This question was addressed by Oskar Klein
in his seminal paper: {\it ``Zur quantenmechanischen begr\"{u}ndung des zweiten
Hauptsatzes der W\"{a}rmelehre''} \cite{klein1}.
He found a fundamentally new way for information to be lost hence  entropy to
increase, special to quantum mechanics.

This result is called Klein's lemma. Let me illustrate his reasoning using a
simple example.
Consider the pure density matrix
\begin{equation}
\rho = \pmatrix{
1 & 0 \cr
0 & 0 \cr
}                                     ,
\end{equation}
with entropy $S=0$. The precise form of the density matrix depends on the
wave-function basis in which it is expressed. For instance, we might choose
another basis such that it becomes
\begin{equation}
\rho ={1 \over \sqrt{2}}\left (\matrix{
1 & 1 \cr
}\right )\otimes {1 \over \sqrt{2}}\left (\matrix{
1 & 1 \cr
}\right ) = \pmatrix{
1/2 & 1/2 \cr
1/2 & 1/2 \cr
}                  .
\end{equation}
Let us now ignore the off-diagonal elements in the new basis. We then get
\begin{equation}
\rho \rightarrow \pmatrix{
1/2 & 0 \cr
0 & 1/2 \cr
}                   .
\end{equation}
If we now attempt to calculate the entropy we find $ S= k \log 2 >0$. If we can
argue, in some specific basis, that the off diagonal terms are irrelevant we
have obtained an increase of entropy.
The main idea is that relevant information can, through some process akin to
the Sto$\beta$zahlansatz of Boltzmann, be transferred into irrelevant quantum
phase information.  Averaging over the phases makes the off-diagonal terms
vanish. As we will see later, this particular way for the entropy to increase
is of fundamental importance.
Hence, I will have reason to come back to Klein's lemma further on.

\section{Ways to find out}

\subsection{Calculations}

To fully understand the physics of quantum black holes, it is necessary to have
a theory of quantum gravity. It has been incredibly difficult to find such a
theory. The only candidate we have today where at least some calculations can
be made is string theory.
Clearly it would be desirable to take a theory of a four dimensional
string (such exist) and investigate what happens when a black hole
forms and evaporates. A full quantum calculation would presumably
teach us a lot. Unfortunately this is beyond our present mathematical ability.
Instead one is, in reality, forced to guess and approximate. And
at the end of the day there are a lot of different opinions and wild
disagreement. Perhaps one should consider a much simplified model which
can be solved exactly? Such models can be found in two dimensions.
Since gravity is renormalizable in two dimensions we can consider models
both with and without strings. Lately, following the work  [11], much attention
has been given to the two dimensional space-time action
\be
S =  \frac{1}{4\pi} \int d^{2}x \sqrt{-g} e^{-2\phi} (R +4(\nabla \phi )^{2} +
\lambda ^{2})    .
\label{verk}
\ee
It can be derived from string theory by demanding world-sheet conformal
invariance
at the quantum level. Only solutions of the equations of motion derived
from the above action obey this requirement.

The expression (\ref{verk}), and the presence of a dilaton, can be
motivated also from a purely field theoretic point of view. Unlike
other dimensions, the pure Einstein action is a topological invariant
in two dimensions, the Euler characteristic. Hence we must complicate
life a bit -- the introduction of the dilaton is natural.

The simplest solution of the equations of motion is the linear dilaton
vacuum given by
$$
ds^{2} = -dt^{2} + dx^{2}
$$
\be
\phi = -\lambda x     .
\ee
Another solution is the black hole \cite{svart} given by
$$
-(1-\frac{M}{\lambda} e^{-2\lambda x} )dt^{2} +\frac{1}{1-\frac{M}{\lambda}
e^{-2\lambda x}} dx^{2}
$$
\be
\phi = -\lambda x    .
\ee
The black hole has been studied extensively. It can be shown to Hawking radiate
with a temperature $T=\frac{\lambda}{2\pi}$. This is most easily seen by going
to Euclidean signature and making sure that there is no conical singularity on
the horizon. In two dimensions it is also possible to use the well developed
techniques of conformal field theory. In particular, it is easy to evaluate the
expectation value of the energy momentum tensor. Naively it is infinite.
However, in what we choose to call the vacuum it must, by definition, be zero.
This is fixed by normal ordering where the infinity is subtracted. The
subtraction depends on which coordinate system is used. In other words,
different observers using different coordinate systems have different opinions
of what the vacuum is. This is the cause of the Unruh effect \cite{unruh}
where accelerated observers see thermal radiation.

Let me now make a small digression to better understand the above and to hear
what Klein might have to say.
Hawking radiation from a black hole and the Unruh radiation are, presumably,
equivalent phenomena related by the equivalence principle. A careless
application of the equivalence principle would, however, imply the existence of
radiation not only for an observer hovering above a black hole, but also above,
let's say, a neutron star. This is however not correct. Such an observer would
not see any radiation. The vacuum state over a not fully collapsed object is
different from that above a black hole due to different boundary conditions
\cite{ginz}.
We understand from this that the vacuum is an object with much structure.
Continuing the historical comparisons it  is very similar to the concept of the
ether of the previous century. Let me quote Klein \cite{klein2}:

\begin{flushleft}
{\it ``... the practical situation of general relativity theory taken together
with our present knowledge regarding matter and the vacuum would rather point
to the assumption that the world as we know it is to be regarded as a weak
excitation of the vacuum state, which state, in spite of its relative
character, may be compared to the absolute space of Newton."}
\end{flushleft}

Let me now come back to the two dimensional black hole!
The energy momentum tensor will necessarily react back on the geometry. Among
other things, the mass of the black hole must decrease! Some of these effects
can be studied by adding a correction to the action and then solve the
modified, semi-classical, equations of motion.
Attempts have then been made to follow the evaporation process towards the end
\cite{suss1a,suss1b,suss1c}. Unfortunately, no definite conclusions can be
drawn. However, some of the more elaborate calculations
\cite{erika,erikb,erikc}, give remarkable hints indicative of the scenarios
that I will discuss in the next section.

Models of strings moving in two dimensional space-times have been studied using
other much more powerful methods; the matrix models. Matrix models represent
the world-sheet of a string through triangulations. These triangulations are
then given by Feynman-diagrams of matrix fields.
For a review and list of references, see [22].  Remarkably, these techniques
allow for  the exact evaluation of correlation functions in the linear dilaton
theory.

There have also been attempts to give a matrix model description of the black
hole \cite{jev,min2,dem}. Clearly it would be wonderful to have such a tool!
Even in these models exact correlation functions can be calculated. In
principle these results should tell us a lot about the information paradox, if
only we had the means to interpret the mathematics.

Let us now leave these mathematical considerations and move into
much less restricted areas.

\subsection{Speculations}

I now want to describe the black-hole information paradox in more detail.
We will see that the key-question is: ``Do you
loose your memory if you travel into a black hole?".
Clearly, this should not be the case according to the equivalence
principle. For a large black hole the tidal forces are weak, even
close to the horizon. There is nothing exceptional, locally, at the point of no
return.
A black-hole explorer could safely travel
through the horizon and into the black hole. How would an observer
remaining outside interpret this journey? If we believe in the unitarity
of quantum mechanics and that all information about what went inside
should be possible to recover just by looking closely (no true
entropy!), we must conclude that information about the infalling
observer (including quantum phases) are contained in the Hawking radiation.

But quantum mechanics strikes back! It is impossible, according to quantum
mechanics, to make perfect copies and keeping the original \cite{zur1}. No
unitary process can duplicate an arbitrary unknown initial state. For some
thoughts on the subject see [27]. So, {\it if} the Hawking radiation really do
contain all information about the matter that formed the black hole, then no
information can pass into the black hole. You would in other words loose your
memory if you attempted to travel into a black hole! Clearly we have run into a
paradox.

Is there a way out? Recently a possible resolution of the paradox has emerged.
This is ``Black Hole Complementarity''. It says that:

\begin{flushleft}
{\it ``You shall not speak about the inside and the outside of a black hole at
the same time. The notebooks of an inner and an outer observer can never be
compared and hence need not agree."}
\end{flushleft}

Black hole complementarity can be realized in two ways, each related to an
uncertainty principle. The first one is {\it  quantum mechanical
complementarity}. It uses the fact that questions about physics far out from
the black hole and questions about physics close to the horizon might be
complimentary in a quantum sense. Mathematically speaking, the corresponding
operators do not commute \cite{thooft}. This is easy to see. It is well known
that the expectation values of the energy-momentum tensor is not very big at
the horizon. This I discussed briefly in the previous subsection. Hence, it is
concluded, the back reaction is not a serious problem and the semi-classical
approach of Hawking basically correct. However, this is the picture appropriate
to an {\it infalling} observer. Let us now consider a measurement far out where
a Hawking photon is detected. Once detected, it is justified to trace it back
in time to see where it came from. Due to the large blueshift, it is found to
have had extremely high energ

y close to the horizon. The back-reaction will be considerable and the standard
picture breaks down. In the next section I will discuss a framework which is
useful in this context.

The second possible reason for complementarity is {\it string theoretical
complementarity}. It comes from the fact that a string looks larger when
redshifted! \cite{suss2a,suss2b,suss2c} The reason is that the size of the
string depends on the cut-off used for its vibrational modes. The higher the
cut-off (i.e. the closer we examine the string) the larger it looks. Hence, an
observer far out looking at a string near the horizon would, through the
redshift, use a smaller cut-off (in the string frame) than an observer
travelling along with the string. Since everything is supposed to be composed
of strings the outer observer's view of the black hole would be blurred. He
would not be able to make accurate statements about the space time structure
near the horizon. Indeed, he might not be able to claim that there is matter
falling into the black hole at all!

\subsection{Dreams}

Let me now discuss the problem from a different point of view.
Is true entropy in any way needed? Does it fit naturally into some part of
physics? Let us consider this question
in the context of the measurement problem of quantum mechanics.
This has been a very popular subject and a starting point for various
claims about quantum mechanics being incomplete.

The heart of the mystery is the collapse of the wave-function. Let us
look at this more closely. There seems to be two different time
evolutions in quantum mechanics. The {\it unitary} evolution between
measurements, and the {\it non-unitary} collapse at a measurement.

The first is described by the Schr\"{o}dinger equation. The collapse, however,
is not.
The collapse proceeds in two steps. The first one involves the disappearance of
superpositions. In other words, the off-diagonal elements of the density matrix
become zero:
\begin{equation}
\rho  = \pmatrix{
\rho _{11} & \rho _{12} \cr
\rho_{21} & \rho _{22} \cr
} \rightarrow \pmatrix{
\rho _{11} & 0 \cr
0 & \rho _{22} \cr    }      .
\label{koll}
\end{equation}
This means that the very quantum mechanical object on the left, describing a
superposition, is transformed into a statistical sum of different possibilities
with probabilities on the right. This is needed in order to make predictions.
The world we directly observe is always classical. We can never even {\it
imagine} anything else. When we read our measurement apparatus we never,
directly, confront superpositions or amplitudes. No one has ever directly seen
a superposition. We only have classical readings of measuring devices which we
interpret.
At most, we need to consider probabilities. The existence of a quantum world
may only be inferred from our measurements. The only questions we can ask are
classical, or, more precisely, based on classical logic. The second step of the
collapse, then, is the realization of a particular alternative according to the
probabilities supplied by the first step.

The standard Copenhagen interpretation of quantum mechanics postulates a
classical world to which we can relate. It would be  desirable to {\it derive}
its existence through more basic means. So, what about  (\ref{koll}) above? It
is clearly impossible to describe such processes within unitary quantum
mechanics. Also, the unitary evolution is certainly deterministic while the
world is not. In fact, this is the main message of quantum mechanics. The
collapse of the wave function seems to be an extremely important part of  the
quantum mechanical world. Hence it would be desirable to give it a more precise
description. Is it a physical process? It is clearly tempting to imagine
non-unitary phenomena responsible not only for the appearance of the classical
world but, perhaps, also associated with black-holes.

However, I will argue that this need not be the case, and I'm sure Oskar Klein
would have agreed. He would have considered, as I have explained earlier, the
vanishing of the off-diagonal terms as being due to the interaction with the
environment. In fact, (\ref{koll}) can be shown to be almost unavoidable in the
physical world. For instance, it can be calculated that the moon is decohered
and put to a classical well defined position through scattering of photons from
the microwave background in just a tiny fraction of a second.

The best way to systematize how decoherence helps in producing the classical
world is the {\it consistent history approach}. This is not the place to give a
detailed explanation, instead I refer to the many excellent review papers which
exist \cite{hart1}.

A common complaint is that it is not correct to think of the mixed density
matrix  as just statistical, even though it {\it looks} that way. When the
environment is included, we {\it still} have superpositions. It is our limited
knowledge that  gives the illusion of a mixed state. It is true that the first
step alone provides us with the means of making statistical predictions, but
the second step is needed to realize a particular outcome. After all, in the
real world just {\it one} thing happens. Otherwise the statistical predictions
would be meaningless. While the first step can be described, as we have seen,
within unitary quantum mechanics, this seems impossible for the second step.
{}From many points of view it is really this second step of the collapse that
is the most  mysterious.

However, it is important to realize that only the first step of the collapse
{\it needs} a physical description. If the first step can be managed, then we
can do without the Copenhagen way of arbitrarily separating the world into
quantum and classical. The theory will do it for us. It will even be possible
to discuss the quantum mechanics of the whole universe as a closed system.
The understanding of decoherence achieved during the last few years solves this
problem.
What about the second step? In this case there is no hindrance in managing {\it
without} a physical description.  Quantum mechanics is  a probabilistic theory.
We should be satisfied when we have obtained the probabilities and not ask for
more. One should not confuse the theoretical {\it representation} of the world
with the physical world itself. Step one of the collapse is fully describable
within the mathematical representation, the second step is not, or at least,
need not be for consistency. The Copenhagen interpretation did not allow even a
description of the first step. This was clearly a limitation. Excluding the
{\it second} step is no limitation however, but rather a success and the way
quantum mechanics avoids determinism.  This point of view has been nicely
explained in [33]. Hence there are no definite reasons to invoke non-unitarity.
It has no role within our {\it representation} of the world.

It is interesting to speculate further on the laws of consistent histories and
the emergence of a classical world. What if there were several {\it different}
possibilities? The same wave function of the universe but with several parallel
classical worlds? This has been discussed in [34]. Well, I can give a
suggestion for why this could be important: ``Black Hole Complementarity''.
According to the quantum mechanical version of black hole complementarity, two
observers, one close to the horizon and the other one far away, will be making
non-commuting observations. As the first observer  is approaching the horizon,
their respective set of consistent histories will gradually become mutually
inconsistent.
This will be true even at a macroscopic level, the two observers  can not even
agree upon the mere existence of the other observer! We find that the
measurement problem of quantum mechanics, when properly understood, hints at a
solution of the information paradox without non-unitarity rather than, as one
naively would expect, gives support for extensions of quantum mechanics.

To conclude, it is hard to find a real need for true entropy.  The measurement
problem has received a, for practical purposes, satisfactory treatment.  We
have still to discover strong reasons to believe that quantum mechanics needs
modification. But, perhaps, some other day we will. After all, black hole
complementarity is just a scenario and we do not know to what extent it is
realized in Nature.

\section{Conclusions}

I have considered three different approaches to the question of true
entropy. The first two focused on black holes. Models of quantum black holes do
exist, at least within string theory. However, the mathematics is formidable.
It is very difficult to draw any definite conclusions.

Another approach is to forget about mathematical rigor and search for
consistent scenarios only. Remarkably, as have been realized recently, there is
an interesting possibility referred to as ``Black Hole Complementarity''. It
suggests that true entropy is avoidable in black hole evaporation.

Then I considered the question: what if there were true entropy, would it be of
any use? To this end I considered the measurement problem of quantum mechanics
but, alas, no application could be found. However, an interesting connection
with black hole complementarity was noted.

To fully appreciate the black hole information paradox and its possible
resolution through black hole complementarity, it is necessary to understand
the measurement problem of quantum mechanics. The connection, as I have argued,
is not that they imply non-unitary physics -- they do not -- instead they both
illuminate the subtlety of quantum mechanics and the emergence of the classical
world.

Oskar Klein identified, more than 60 years ago, one of the deepest problems of
physics.
We are still waiting for the full answer.

\section{Acknowledgements}

I would like to thank E. Verlinde for  stimulating discussions.

\end{document}